\documentclass[%
 aip,
 amsmath,amssymb,
 reprint,%
]{revtex4-1}

\usepackage{graphicx}% Include figure files
\usepackage{dcolumn}% Align table columns on decimal point
\usepackage{bm}% bold math
\usepackage[utf8]{inputenc}
\usepackage[T1]{fontenc}
\usepackage{mathptmx}

\newcommand{\la}{\lambda}
\newcommand{\ga}{\gamma}

\newcommand{\om}{\omega}

\newcommand{\vphi}{\varphi}

\newcommand{\prt}{\partial}

\newcommand{\sn}{\mathrm{sn}}

\newcommand {\e} {\mathrm{e}}

\begin{document}

%\preprint{AIP/123-QED}

\title{Evolution of instability fronts in sine-Gordon equation dynamics }

\author{ A. M. Kamchatnov}

\affiliation{Institute of Spectroscopy,
Russian Academy of Sciences, Troitsk, Moscow, 108840, Russia}

\email{kamch@isan.troitsk.ru.}

\date{\today}

\begin{abstract}
Solutions of the Whitham modulation equations for one-phase periodic waves obeying the sine-Gordon equation 
are found that describe the evolution of an oscillatory region behind an instability front propagating into 
the instability region. A simple self-similar solution describes the whole region between two fronts of 
instability resulting from a localized initial disturbance in the unstable state. Another hodograph 
solution represents typical waves close to the instability fronts. This theory generalizes the approach 
used previously for systems obeying the nonlinear Schr\"{o}dinger equation.
\end{abstract}


% 47.35.Jk 	Wave breaking
% 02.30.Ik 	Integrable systems
% 02.30.Jr 	Partial differential equations
% 47.35.Fg 	Solitary waves

\maketitle

\begin{quotation}
It is well known that the Whitham modulation theory provides quite a general approach to studying 
the stability of nonlinear waves. In case of stable waves, the Whitham method forms the basis for 
the theory of dispersive shock waves. However, this method applied to modulationally unstable 
systems turned out less informative because of the extremely high sensitivity of solutions to 
small non-analytic changes in initial conditions. In spite of that, the method allows one to make 
some quite general statements about the evolution of modulationally unstable systems. In particular, 
a local disturbance of an unstable state leads to the formation of an instability wave propagating 
into the region of instability with some velocity, and the value of this velocity is determined by 
the properties of the system. This problem is well studied for nonlinear diffusive systems, and in 
this paper we use the Whitham approach for studying conservative modulationally unstable systems. 
Since the modulation instability is the result of the interplay of nonlinear and dispersive 
properties of the system, it is natural to suppose that these properties lead to the formation of 
a region of nonlinear oscillation that expands with time. Therefore, after long enough time of 
evolution, we have a long region of relatively fast oscillations, so we can average over these 
oscillations and study the slow evolution of the modulation parameters. In this paper, we apply 
this idea to the celebrated sine-Gordon equation, which has unstable solutions known as a false 
vacuum. Then a local disturbance evolves into a long region of oscillations, and equations for 
evolution of the averaged modulation parameters were derived long ago by Whitham. We have studied 
two types of solutions of these equations. First, we obtained a simple self-similar solution, 
which describes qualitatively the whole region of oscillations. Second, we found particular 
solutions that illustrate the propagation of the instability front into the unstable state. 
In both cases the velocity of the instability front is close to unity in standard non-dimensional 
units.
\end{quotation}

\section{Introduction}\label{intro}

Unstable extended physical systems are omnipresent in nature. Elementary physics provides such
simple examples as a metastable supersaturated vapor where tiny pieces of dust or ions trigger
the formation of liquid drops. In some situations, the transition from an unstable state to a
stable one takes place through the propagation of a front of crystallization, the propagation
of the flame, or the propagation of a mutant gene in a population (see, e.g., \cite{scott-07}).
The velocity of such a front propagation is determined by the intrinsic properties of the system,
and it was studied in a number of papers (see, e.g., \cite{kpp-37,dl-83,kn-87,cgk-94,vanSaar-03}
and references therein).

It is remarkable that the propagation of instability fronts in conservative modulationally
unstable systems is described by a similar theory. For example, a plane wave whose evolution
obeys the nonlinear Schr\"{o}dinger (NLS) equation is unstable with respect to disintegration
into wave packets. As was found first numerically \cite{karpman-67,kk-68,karpman}, a small
localized perturbation leads to the formation of a region of nonlinear oscillations, and this
region gradually expands along the unstable plane wave. If we represent this region of nonlinear
oscillations as a modulated periodic wave, then at a large enough time of evolution the Whitham
theory \cite{whitham-65,whitham} can be applied, and this method provides an important result
that the small-amplitude edges of the region of nonlinear oscillations propagate with the
minimal group velocity of the stable mode \cite{kamch-92,egkk-93,bk-94}. A similar conclusion
was made for the propagation of the instability fronts of two-dimensional dark solitons generated
by a flow of a Bose-Einstein condensate past an obstacle \cite{kt-88,egk-06,kp-08,kk-11} with
agreement of the experimental observations of Refs.~\cite{amo-11,grosso-11}.

However, there are different scenarios of instability fronts propagation. For example, let a
string lie at the top of the local maximum of the potential $U(\vphi)$ at $\vphi=\vphi_{max}$,
where $\vphi$ is the wave variable whose dynamics obeys the generalized Klein-Gordon equation
\begin{equation}\label{eq1}
  \vphi_{tt}-\vphi_{xx}+U'(\vphi)=0,\quad U'=\frac{dU}{d\vphi},\quad U(0)=0.
\end{equation}
If $U(\vphi)=1-\cos\vphi$, then we arrive at the celebrated sine-Gordon equation. In this case, 
the state $\vphi=0$ is stable, and the state $\vphi=\vphi_{max}=\pi$ is obviously unstable. If we 
disturb this unstable state locally in the vicinity of the point $x=0$, then the left- and 
right-propagating waves along negative and positive directions of the $x$-axis will form an 
oscillatory structure between two instability fronts. Now there are no stable propagating 
modes along the state $\vphi=\vphi_{max}=\pi$, so there is no local maximum of the group velocity. 
Thus, we arrive at the problem of the description of the evolution of the oscillatory structure 
and, in particular, of finding the velocity of the propagation of the instability fronts. This 
paper is devoted to the discussion of these problems.

\section{Periodic solutions and modulation equations}

We are going to represent an instability wave as a modulated periodic solution of the corresponding 
nonlinear wave equation. So, first of all, we have to get the periodic solutions and obtain the 
modulation equations. Actually, these two problems were discussed for the sine-Gordon equation 
in a number of papers (see, e.g., Refs.~\cite{akns-73,takhtajan-74,kk-76,ks-91,gn-03,cmkw}). 
However, for our aim, it is more convenient to consider first the generalized Klein-Gordon equation 
(\ref{eq1}) and then to specialize the results for the sine-Gordon equations.

As is well known, Eq.~(\ref{eq1}) has traveling wave solutions $\vphi=\vphi(\xi)$, $\xi=x-Vt$, where $\vphi(\xi)$
is defined implicitly by the equation
\begin{equation}\label{eq2}
  \xi-\xi_0=\sqrt{\frac{V^2-1}{2}}\int_{\vphi_0}^{\vphi}\frac{d\vphi}{\sqrt{A-U(\vphi)}},
\end{equation}
so that $V$ and the integration constant $A$ are parameters, $\vphi(\xi_0)=\vphi_0$, and the variable
$\vphi$ oscillates between two roots of the equation $A-U(\vphi)=0$ in the positivity interval of
this expression. Following Whitham \cite{whitham-65,whitham}, we define the function
\begin{equation}\label{eq3}
\begin{split}
  W(V,A)&=\sqrt{2(V^2-1)}\oint \sqrt{A-U(\vphi)}\,d\vphi\\
  &\equiv\sqrt{V^2-1}\cdot G(A),
  \end{split}
\end{equation}
where the integration is taken along a contour around this positivity interval. Then the wavelength
of the above solution is given by the expression
\begin{equation}\label{eq4}
  L=\frac{\prt W}{\prt A}=\sqrt{V^2-1}\cdot G'(A).
\end{equation}
We define the wave number as $k=1/L$, so that
$k^2(V^2-1)=(G')^{-2}$, and obtain for the frequency $\om=kV$ the dispersion relation
\begin{equation}\label{eq5}
  \om^2=k^2+(G'(A))^{-2},
\end{equation}
which depends essentially on the amplitude $A$. The group velocity is defined as
\begin{equation}\label{eq6}
  v=\left(\frac{\prt\om}{\prt k}\right)_A=\frac{k}{\om}=\frac{1}{V}.
\end{equation}

In a modulated wave the parameters $V$ and $A$ become slow functions of $x$ and $t$, and their evolution
obeys the Whitham modulation equations \cite{whitham-65,whitham} which in our notation can be
written in the form
\begin{equation}\label{eq7}
  \begin{split}
  & \left(\frac{kV}{V^2-1}+A\right)_t+\left(\frac{kVW}{V^2-1}\right)_x=0,\\
  & \left(\frac{kVW}{V^2-1}\right)_t+\left(\frac{kV^2W}{V^2-1}-A\right)_x=0.
  \end{split}
\end{equation}
It is convenient to exclude $W$ and $V$ with
the use of Eqs.~(\ref{eq3}), (\ref{eq6}) and then we get
\begin{equation}\label{eq8}
  \begin{split}
  & \left(\frac{G/G'}{1-v^2}+A-\frac{G}{G'}\right)_t+\left(\frac{(G/G')v}{1-v^2}\right)_x=0,\\
  & \left(\frac{(G/G')v}{1-v^2}\right)_t+\left(\frac{(G/G')v^2}{1-v^2}-A+\frac{G}{G'}\right)_x=0.
  \end{split}
\end{equation}
Linearization of these equations with respect to small deviations from constant values
$A$ and $v$ yields the characteristic velocities
\begin{equation}\label{eq9}
  v_{\pm}=\frac{v\pm c}{1\pm vc},
\end{equation}
where $c$ is defined by the expression
\begin{equation}\label{eq10}
  c^2=-\frac{GG^{\prime\prime}}{(G')^2}.
\end{equation}
If its right-hand side is positive, then formulas (\ref{eq9}) have simple physical sense: they
give velocities of propagation of a sound signal with the speed $c$ upstream or downstream the
flow of a ``fluid'' moving with velocity $v$, so that in the laboratory reference frame the
velocity of the signal is equal to the relativistic sum of these two velocities. 
It is worth noticing that the Whitham equations (\ref{eq8}) can be cast to the diagonal
Riemann form
\begin{equation}\label{eq11}
  \frac{\prt r_{\pm}}{\prt t}+v_{\pm}\frac{\prt r_{\pm}}{\prt x}=0
\end{equation}
for the variables
\begin{equation}\label{eq12}
  r_{\pm}=\frac12\ln\frac{1+v}{1-v}\mp\int^A\frac{cG'}{G}\,dA,
\end{equation}
called Riemann invariants.

Eq.~(\ref{eq1}) is relativistically invariant, and this suggests that the modulation equations
(\ref{eq8}) can be interpreted as equations of relativistic hydrodynamics \cite{maslov-69}.
To show this explicitly, we recall that the equations of relativistic hydrodynamics follow
from the energy-momentum conservation law in a relativistic flow (see\cite{LL6}),
\begin{equation}\label{eq13}
  \frac{\prt T^{00}}{\prt t}+\frac{\prt T^{10}}{\prt x}=0,\quad
  \frac{\prt T^{10}}{\prt t}+\frac{\prt T^{11}}{\prt x}=0,
\end{equation}
where
\begin{equation}\label{eq14}
  T^{ij}=wu^iu^j-pg^{ij},\quad i,j=1,2,
\end{equation}
is the energy-momentum tensor in two-dimensional Minkowski space with the metric tensor
\begin{equation}\label{eq15}
  g^{ij}=\left(
           \begin{array}{cc}
             1 & 0 \\
             0 & -1 \\
           \end{array}
         \right),
\end{equation}
$w=e+p$ is the enthalpy density, $e$ is the energy density, $p$ is the pressure, and
$u^i$ is a two-dimensional vector of ``4-velocity''. To identify the Whitham equations (\ref{eq8})
with Eqs.~(\ref{eq13}), we introduce in a usual way the 4-vector  $u^i$,
\begin{equation}\label{eq16}
  u^0=\frac{1}{\sqrt{1-v^2}},\qquad u^1=\frac{v}{\sqrt{1-v^2}},
\end{equation}
and then it is easy to see that equations (\ref{eq8}) and (\ref{eq13}) coincide with each other if
\begin{equation}\label{eq17}
  \begin{split}
  & T^{00}=\frac{w}{1-v^2}-p=\frac{G/(2G')}{1-v^2}+\frac{A}2-\frac{G}{2G'},\\
  & T^{10}=T^{01}=\frac{wv}{1-v^2}=\frac{(G/(2G'))v}{1-v^2},\\
  & T^{11}=\frac{wv^2}{1-v^2}+p=\frac{G/(2G')}{1-v^2}-\frac{A}2+\frac{G}{2G'},
  \end{split}
\end{equation}
where we divided Eqs.~(\ref{eq8}) by 2 for further convenience. Consequently we get the
relationships of the wave amplitude $A$ with thermodynamical functions
\begin{equation}\label{eq18}
  e=\frac{A}{2},\quad p=\frac{G}{2G'}-\frac{A}2,\quad w=e+p=\frac{G}{2G'},
\end{equation}
of effective matter whose dynamics obeys equations (\ref{eq13}).
It is important that the pressure $p$ only depends on the energy density $e$. This means that the
mass of particles in the effective matter is negligibly small.
The expression (\ref{eq10}) for the sound velocity can be written in the standard form
\begin{equation}\label{eq19}
  c^2=\frac{dp}{de}.
\end{equation}
The temperature $T$ and the entropy density $\sigma$ of the effective matter can be defined in
the following way. The chemical potential of a gas of massless particles equals to zero, so
the enthalpy is given by the formula $w=T\sigma$, consequently the relationship
$dw=Td\sigma+dp=d(T\sigma)$ gives $dp=\sigma dT$
(see \cite{LL5}). Hence, the squared sound velocity can be written in two forms: from (\ref{eq10})
and (\ref{eq18}) we get
$$
c^2=-w\frac{d^2G/de^2}{dG/de},
$$
whereas Eq.~(\ref{eq19}) with account of $dp=\sigma dT=(w/T)dT$ yields
$$
c^2=\frac{w}{T}\frac{dT}{de}.
$$
Comparison of these two expressions gives the formulas
\begin{equation}\label{eq20}
  T=\gamma\cdot\left(\frac{dG}{de}\right)^{-1},\qquad \sigma=\frac1{\gamma}\cdot G(e)
\end{equation}
with the same numerical factor $\gamma$ in both formulas. Then we obtain the relation
\begin{equation}\label{eq20b}
  d\sigma=\frac1{\ga}\frac{dG}{de}de=\frac{de}{T},
\end{equation}
which agrees with the standard thermodynamical definition of the entropy.

The hydrodynamic equations get especially simple form in the variables $T,\sigma, u^i$.
We notice that the formulas for the wave vector $k=1/L$ and the frequency $\om=kV=k/v$
are transformed to
\begin{equation}\label{eq21}
\begin{split}
  & k=2u^1\left(\frac{dG}{de}\right)^{-1}=\frac2{\ga}u^1T,\\
  & \om=k\frac{u^0}{u^1}=\frac2{\ga}u^0T.
  \end{split}
\end{equation}
Consequently the conservation of the number of waves law, which follows from Eqs.~(\ref{eq7})
(see \cite{whitham-65,whitham})
\begin{equation}\label{eq22}
  k_t+\om_x=0,
\end{equation}
transforms to
\begin{equation}\label{eq23}
  (u^1T)_t+(u^0T)_x=0.
\end{equation}
It is worth noticing that this relation was obtained by I.~M.~Khalatnikov \cite{khal-54}
from Eqs.~(\ref{eq13}) for any one-dimensional relativistic flow what proves that it is
potential. In Whitham's theory Eq.~(\ref{eq22}) follows from definition of the wave vector
and the frequency as the phase derivatives, $k=\theta_x,\om=-\theta_t$. We see that both
pictures, Whitham's modulation and relativistic hydrodynamical ones, agree mathematically
and differ only in notation and physical meaning of variables.

One more equation we obtain from the formula
\begin{equation}\label{eq24}
  \frac{\prt(wu^i)}{\prt x^i}-u^i\frac{\prt p}{\prt x^i}=0,
\end{equation}
which is a consequence of Eqs.~(\ref{eq13}), (\ref{eq14}) (see Eq.~(134.5) and problem 2
in Ref.~\cite{LL6}, \S134). Substitution of $w=T\sigma$, $dp=\sigma dT$ gives at once
the equation
\begin{equation}\label{eq25}
  \frac{\prt(\sigma u^0)}{\prt t}+\frac{\prt(\sigma u^1)}{\prt x}=0,
\end{equation}
which means conservation of entropy, i.e. the flow is adiabatical.

\begin{figure}[t]
    \centering
    \includegraphics[width=8cm]{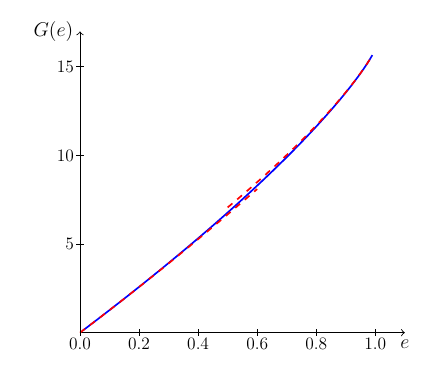}
    \caption{Plot of the function $G(e)$ defined in Eq.~(\ref{eq27}). Red dashed lines
    correspond to the asymptotic expressions (\ref{eq27a}) and (\ref{eq27b}).}
    \label{fig1}
\end{figure}

Let us specify these relation for the sine-Gordon equation when we have in Eq.~(\ref{eq1})
\begin{equation}\label{eq26}
  U'(\vphi)=\sin\vphi,\qquad U(\vphi)=1-\cos\vphi.
\end{equation}
Then the integral in Eq.~(\ref{eq2}) reduces to the elliptic one of the 1st kind and its
inversion yields the periodic solution in explicit form
\begin{equation}\label{eq26a}
\begin{split}
  \vphi&=2\arcsin\left[\sqrt{e}\,\,\sn\left(\frac{\xi-\xi_0}{\sqrt{V^2-1}},e\right)\right]=\\
  &=2\arcsin\left[\sqrt{e}\,\,\sn\left(\frac{v(x-x_0)-t}{\sqrt{1-v^2}},e\right)\right].
  \end{split}
\end{equation}
In the limit $e\to1$ this solution converts into the known kink solution of the sine-Gordon equation
\begin{equation}\label{eq26b}
  \vphi=2\arcsin\left[\tanh\left(\frac{v(x-x_0)-t}{\sqrt{1-v^2}}\right)\right],
\end{equation}
so that $\vphi=-\pi$ as $x\to-\infty$ and $\vphi=\pi$ as $x\to+\infty$. The integral in
Eq.~(\ref{eq3}) can be reduced to the elliptic ones, so we get
\begin{equation}\label{eq27}
\begin{split}
  G(e)&=2\sqrt{2}\int_{-\vphi_m}^{\vphi_m}\sqrt{2e-1+\cos\vphi}\,d\vphi=\\
  &=16\{E(e)-(1-e)K(e)\},
  \end{split}
\end{equation}
where $K(e),E(e)$ are the complete elliptic integrals of the 1st and 2nd kind, respectively,
defined here according to the handbook \cite{as-79}; $\pm\vphi_m$ are the roots of the integrand
function, so $\vphi_m$ is the amplitude of oscillations:
\begin{equation}\label{eq27c}
  \vphi_m=2\arcsin\sqrt{e}=\arccos(1-2e).
\end{equation}
The plot of this function is shown in Fig.~\ref{fig1}. For small $e$ it is approximated as
\begin{equation}\label{eq27a}
  G(e)\approx 4\pi e+\frac{\pi}{2}e^2+\ldots,\qquad e\ll1,
\end{equation}
and for $e$ close to unity as
\begin{equation}\label{eq27b}
  G(e)\approx 16-4(1-e)\left(\ln\frac{16}{1-e}+1\right)+\ldots,\quad 1-e\ll1.
\end{equation}
Both asymptotic expressions are shown in Fig.~\ref{fig1} by the red dashed lines and they
provide good enough approximations even for $e\sim0.5$.

We choose for the factor $\ga$ in Eqs.~(\ref{eq20}) the value $\ga=8$ and then the formulas
\begin{equation}\label{eq28}
  \frac{dK}{de}=\frac{E-(1-e)K}{2e(1-e)},\qquad \frac{dE}{de}=\frac{E-K}{2e}
\end{equation}
for derivatives of the elliptic integrals yield the expressions for the thermodynamic functions
\begin{equation}\label{eq29}
  T=\frac1{K},\quad\sigma=4e(1-e)\frac{dK}{de}=2[E-(1-e)K].
\end{equation}
Equation of state $p=p(e)$ has the form
\begin{equation}\label{eq30}
  p=2\left(\frac{E(e)}{K(e)}-1\right)+e.
\end{equation}
The squared sound velocity equals to
\begin{equation}\label{eq31}
  c^2=-4e(1-e)\left(\frac1K\frac{dK}{de}\right)^2
\end{equation}
so the sound velocity is imaginary in the region $0\leq e<1$ of existence of periodic solutions,
that is they describe modulationally unstable waves. Correspondingly, the Riemann invariants
(\ref{eq12}) are complex:
\begin{equation}\label{eq32}
  r_{\pm}=\frac12\ln\frac{1+v}{1-v}\mp i\arcsin\sqrt{e}.
\end{equation}

\section{Self-similar solution of modulation equations}

It is remarkable that the modulation equations in the general form (\ref{eq8}) admit a very
simple self-similar solution that has very clear physical meaning. Indeed, let us suppose that
the initial disturbance of an unstable state is localized in vicinity of the point $x=0$.
Then after long enough time of evolution of the instability wave we will only have small
oscillations around a new stable state $\vphi=\vphi_{max}$ in a large region around this point.
Energy and pressure in the ``effective fluid'' here are small, that is it flows according to
the hydrodynamic-like modulation equations by inertia, so we have here
\begin{equation}\label{eq32}
  v=\frac{x}{t}.
\end{equation}
Substitution of this flow velocity distribution into Eqs.~(\ref{eq8}) yields the system
\begin{equation}\label{eq33}
  \begin{split}
  & \left(\frac{x^2}{t^2-x^2}\frac{G}{G'}+A\right)_t+\left(\frac{xt}{t^2-x^2}\frac{G}{G'}\right)_x=0,\\
  & \left(\frac{xt}{t^2-x^2}\frac{G}{G'}\right)_t+\left(\frac{x^2}{t^2-x^2}\frac{G}{G'}-A\right)_x=0.
  \end{split}
\end{equation}
Since $G$ is a function of $A$, this is actually the system for $A=A(x,t)$ or $G=G(x,t)$. The
relativistic invariance of Eqs.~(\ref{eq8}) or (\ref{eq33}) suggests that at asymptotically large time
these functions can only depend on the relativistic invariant
\begin{equation}\label{eq34}
  s=\sqrt{t^2-x^2},
\end{equation}
and then both Eqs.~(\ref{eq33}) reduce to the same equation
\begin{equation}\nonumber
  \frac{G}{G'}+\frac{dA}{ds}=0\quad\text{or}\quad \frac{d\ln G}{ds}=-1
\end{equation}
with obvious solution
\begin{equation}\label{eq36}
  G(e)=\frac{C}{s}=\frac{C}{\sqrt{t^2-x^2}},
\end{equation}
where $C$ is an integration constant. The applicability condition of the Whitham modulation theory
demands that the integration constant should take such values that the relevant intervals of $x$
contain a large number of oscillations. We are interested in asymptotically large values of $t$,
so the acceptable interval of $x$, $|x|\leq(t^2-C^2/16^2)^{1/2}$, is much greater than the typical
wavelength $L\sim1$ in the wave structure for $C\sim1$. Therefore, to be definite, we will chose
$C=1$ in what follows. The edges of the structure $x_{\pm}=\pm(t^2-1/16^2)^{1/2}$ propagate with
velocities close to unity at $t\gtrsim1$.

\begin{figure}[t]
    \centering
    \includegraphics[width=8cm]{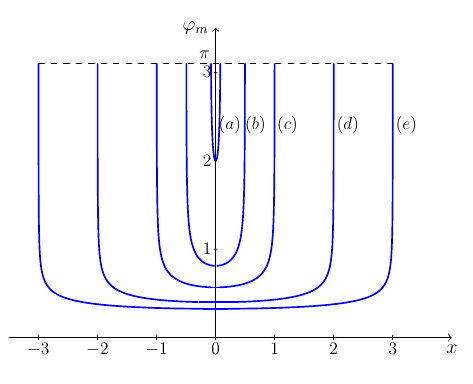}
    \caption{Envelopes of amplitudes $\vphi_m$ for the sine-Gordon instability wave at different moments
    of time: (a) $t=0.1$; (b) $t=0.5$; (c) $t=1$; (d) $t=2$; (e) $t=3$. The dashed line corresponds to the
    maximal amplitude $\vphi_{max}=\pi$.}
    \label{fig2}
\end{figure}

\begin{figure}[t]
    \centering
    \includegraphics[width=8cm]{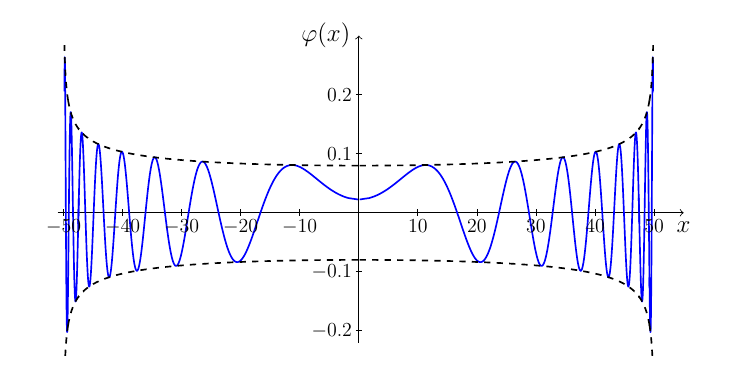}
    \caption{Profile of the instability waves at $t=50$ in the sine-Gordon model.
    The modulated wave (\ref{eq41b}) is
    depicted by a blue line and the enveloped $\pm\vphi_m$ (see Eq.~(\ref{eq27c}) are shown by
    dashed lines.}
    \label{fig3}
\end{figure}

Now let us specify this solution for the case of the sine-Gordon equation when $G(e)$ is given by
Eq.~(\ref{eq27}). Then the formulas
\begin{equation}\label{eq41}
  x(e)=\pm\sqrt{t^2-G^{-2}(e)},\quad \vphi_m(e)=\arccos(1-2e)
\end{equation}
yield in a parametric form the dependence of the envelope function of the amplitude $\vphi_m$ on $x$
for any moment of time $t$, where $e$ plays the role of the parameter. The typical plots are shown
in Fig.~\ref{fig2} (we have chosen small values of $t$ to make evident the evolution of the profile).
As one can see, the instability region expands with time, and its fronts at the edges move
(in the Whitham theory approximation) with the `light' velocity $v=1$.
In the vicinity of the edge points the wave consists of `trains' of kinks with
the amplitudes close to the maximal value $\vphi_m=\pi$. Substitution of Eq.~(\ref{eq32}) into
Eq.~(\ref{eq26a}) yields with account of Eq.~(\ref{eq36})
\begin{equation}\label{eq41b}
\begin{split}
  \vphi= - 2\arcsin\left[\sqrt{e}\,\,\sn\left(G^{-1}(e),e\right)\right],
  \end{split}
\end{equation}
so this equation together with Eq.~(\ref{eq41}) gives the distribution of $\vphi$ on $x$ at some fixed
moment of time $t$. A typical profile is shown in
Fig.~\ref{fig3} for large enough time $t=50$ when we get many oscillations even with $C=1$ in
Eq.~(\ref{eq36}). The dashed lines show the envelopes of the amplitudes $\pm\vphi_m$ calculated
according to Eq.~(\ref{eq27c}).

\begin{figure}[t]
    \centering
    \includegraphics[width=8cm]{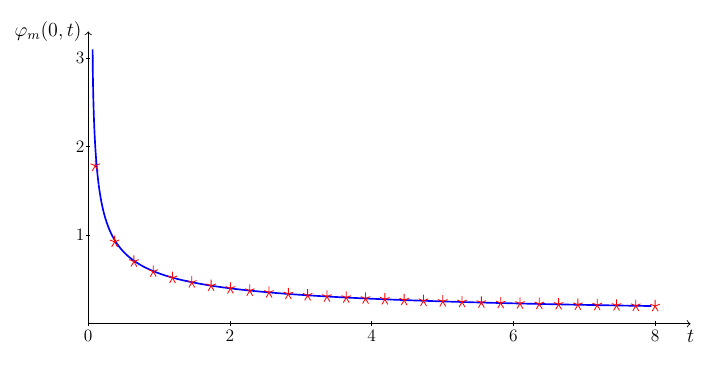}
    \caption{Dependence of the amplitude $\vphi_m(0,t)$ at the center of the expanding wave structure on time $t$.
    Stars correspond to the asymptotic formula (\ref{eq43}), so it provides a good enough approximation
    for $t\gtrsim0.5$.}
    \label{fig4}
\end{figure}

The amplitude decreases at the center of the wave structure which becomes here a wave of oscillations around
the stable state $\vphi=0$. This decrease of the amplitude at the center is given implicitly by the formulas
\begin{equation}\label{eq42}
  G(e)=\frac1t,\quad \vphi_m(e)=\arccos(1-2e)
\end{equation}
and its plot is shown in Fig.~\ref{fig4}. For asymptotically large $t$ (small $e\ll1$) we get
\begin{equation}\label{eq43}
  \vphi_m\approx\frac{1}{\sqrt{\pi t}},
\end{equation}
and this asymptotic formula provides a very good approximation even for relatively large amplitudes $\vphi_m$
(the stars in Fig.~\ref{fig4} correspond to this formula).

On one hand, the solution (\ref{eq36}) is very universal in the sense that it is fulfilled for a wide class
of the nonlinearity potentials $U(\vphi)$ in Eq.~(\ref{eq1}); on the other hand, it is very particular,
since it assumes the motion of the effective relativistic ``fluid'' by inertia according to Eq.~(\ref{eq32}).
Such a motion can be expected for asymptotically large times when a considerable part of the ``matter''
distribution corresponds to a low density, so that the ``pressure'' becomes ineffective. In the next section
we will study the formation of such a region for a large class of solutions of the sine-Gordon equation.

\section{Hodograph transform}

A wide class of solutions of the Whitham modulation equations can be obtained be means of the hodograph transform.
To this end, it is convenient to introduce `rapidity' $y$ instead of velocity $v$
according to the relations
\begin{equation}\label{eq45}
  u^0=\cosh y,\qquad u^1=\sinh y,\qquad v=\tanh y,
\end{equation}
and pass to the `light-cone' variables
\begin{equation}\label{eq46}
x_-=t-x,\qquad x_+=t+x.
\end{equation}
Then Eq.~(\ref{eq23}) takes the form
\begin{equation}\label{eq47}
  \frac{\prt}{\prt x_-}\left(\frac{\e^{-y}}{K(e)}\right)-
  \frac{\prt}{\prt x_+}\left(\frac{\e^{y}}{K(e)}\right)=0,
\end{equation}
and it is satisfied if
\begin{equation}\label{eq48}
  \frac{\e^{-y}}{K(e)}=\frac{\prt\phi}{\prt x_-},\qquad
  \frac{\e^{y}}{K(e)}=\frac{\prt\phi}{\prt x_+}
\end{equation}
for some potential $\phi=\phi(x_-,x_+)$. Now, following Khalatnikov \cite{khal-54}, 
we pass to the potential $\mathcal{W}=\mathcal{W}(e,y)$,
\begin{equation}\label{eq49}
  \mathcal{W}=\phi-K^{-1}\e^yx_--K^{-1}\e^{-y}x_+,
\end{equation}
so that
\begin{equation}\label{eq50}
\begin{split}
  d\mathcal{W}=&-\frac{dK^{-1}}{de}(\e^yx_-+\e^{-y}x_+)de-\\
  &-\frac1K(\e^yx_--\e^{-y}x_+)dy,
  \end{split}
\end{equation}
and, hence,
\begin{equation}\label{eq51}
\begin{split}
  & x_-=-\frac{\e^{-y}}{2}\left(\frac{1}{dK^{-1}/de}\frac{\prt\mathcal{W}}{\prt e}
  +K\frac{\prt\mathcal{W}}{\prt y}\right),\\
  & x_+=-\frac{\e^{y}}{2}\left(\frac{1}{dK^{-1}/de}\frac{\prt\mathcal{W}}{\prt e}
  -K\frac{\prt\mathcal{W}}{\prt y}\right).
  \end{split}
\end{equation}
These formulas perform actually the hodograph transform: if the function $\mathcal{W}=\mathcal{W}(e,y)$
is known, then they give the dependence of $e$ and $y$ on the variables (\ref{eq46}) and,
consequently, on $x$ and $t$.

Equation (\ref{eq25}) after substitutions of (\ref{eq45}), (\ref{eq46}) transforms to
\begin{equation}\label{eq52}
\begin{split}
  &\frac{\prt(\sigma \e^{-y})}{\prt x_-}+\frac{\prt(\sigma \e^y)}{\prt x_+}=\\
 &= \frac{\prt(\sigma \e^{-y},x_+)}{\prt (x_-,x_+)}+\frac{\prt(x_-,\sigma \e^y)}{\prt(x_-, x_+)}=0.
  \end{split}
\end{equation}
Multiplying it by the Jacobian $\prt(x_-,x_+)/\prt(e,y)$, we obtain after simple transformations
with account of Eqs.~(\ref{eq29}) and $d\sigma/de=1/T=K(e)$ the equation for $\mathcal{W}$:
\begin{equation}\label{eq53}
  \frac{\prt}{\prt e}\left[e(1-e)K^2\frac{\prt\mathcal{W}}{\prt e}\right]+
  \frac{K^2}{4}\frac{\prt^2\mathcal{W}}{\prt y^2}=0
\end{equation}
or
\begin{equation}\label{eq54}
  e(1-e)\frac{\prt^2\mathcal{W}}{\prt e^2}+
  \left(\frac{E(e)}{K(e)}-e\right)\frac{\prt\mathcal{W}}{\prt e}
  +\frac14\frac{\prt^2\mathcal{W}}{\prt y^2}=0.
\end{equation}
We can get rid of the elliptic integrals by means of the replacement
\begin{equation}\label{eq54}
  \mathcal{W}(e,y)=\frac{\Phi(e,y)}{K(e)},
\end{equation}
so we obtain
\begin{equation}\label{eq55}
  \frac{\prt}{\prt e}\left[e(1-e)\frac{\prt\Phi}{\prt e}\right]+
  \frac{1}{4}\frac{\prt^2\Phi}{\prt y^2}=\frac14{\Phi}
\end{equation}
or
\begin{equation}\label{eq56}
  e(1-e)\frac{\prt^2\Phi}{\prt e^2}+
  \left(1-2e\right)\frac{\prt\Phi}{\prt e}
  +\frac14\frac{\prt^2\Phi}{\prt y^2}=\frac14{\Phi}.
\end{equation}

These are linear PDEs of elliptic type and they replace the Euler-Poisson equation appearing in
applications of the hodograph transform to the gas dynamics equations (see, e.g., \cite{LL6}).
In modulationally unstable dispersionless dynamics of the NLS equation theory the
analogous equation is the two-dimensional Laplace equation written in polar coordinates
(see, e.g., \cite{lighthill-1965,hayes-73,gs-70}). It is worth mentioning that for small
$e\ll1$ Eq.~(\ref{eq56}) takes after substitution $r=\sqrt{e}$ the form
\begin{equation}\label{eq57}
  \frac{\prt^2\mathcal{F}}{\prt r^2}+
 \frac{1}{r}\frac{\prt\mathcal{F}}{\prt r}
  +\frac{\prt^2\mathcal{F}}{\prt y^2}=\mathcal{F}
\end{equation}
of the Poisson equation instead of the Laplace equation, though the Riemann invariants
(\ref{eq32}) $r_{\pm}=y\mp \sqrt{e}$ coincide with those for the dispersionless limit 
of the NLS equation theory with $y$ playing the role of the flow velocity.

Particular solutions of the linear equations (\ref{eq55}) or (\ref{eq56}) yield some
typical modulated wave patterns which can be formed in the sine-Gordon dynamics. We
will consider here a typical example illustrating propagation of an instability front.

\section{Propagation of the instability front}

We will consider a simple case when the variables in Eq.~(\ref{eq56}) are separated
in the following way:
\begin{equation}\label{eq58}
  \Phi(e,y)=F(e)\exp(\pm i\la y),
\end{equation}
so that $F(e)$ satisfies the hypergeometric equation
\begin{equation}\label{eq59}
  e(1-e)\frac{d^2F}{d e^2}+ (1-2e)\frac{dF}{d e}
  +\frac{\la^2-1}4F=0,
\end{equation}
with the solution $F=F((1+\la)/2,(1-\la)/2,1;e)$ (see, e.g., \cite{as-79,ww-27}).
The substitution $e=(1-z)/2$ casts it to the Legendre equation
\begin{equation}\label{eq60}
  (1-z^2)\frac{d^2F}{dz^2}-2z\frac{dF}{dz}+n(n+1)F=0,\quad n=\frac{\la-1}{2},
\end{equation}
For integer $n=0,1,2,\ldots$ it has solutions in the form of Legendre polynomials
$P_n(z)=P_n(1-2e)$ and Legendre functions of 2nd kind $Q_n(z)=Q_n(1-2e)$.

\begin{figure}
\includegraphics[width=8cm]{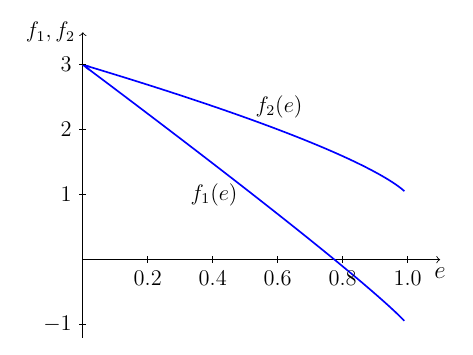}
\caption{Plots of functions $f_1(e)$ and $f_2(e)$ defined by Eqs.~(\ref{eq63}).}
\label{fig5}
\end{figure}

To be definite, we take as an illustrative example the solution with $F=-P_1(z)=2e-1$, 
when $n=1$ and $\la=3$:
\begin{equation}\label{eq61}
  \mathcal{W}(e,y)=\frac{2e-1}{K(e)}\e^{3y}.
\end{equation}
Its substitution into Eqs.~(\ref{eq51}) gives
\begin{equation}\label{eq62}
\begin{split}
  &x_-=t-x=2f_1(e)\e^{2y},\\
  &x_+=t+x=f_2(e)\e^{4y},
  \end{split}
\end{equation}
where
\begin{equation}\label{eq63}
  \begin{split}
  & f_1(e)=\frac{(1-2e)E(e)-(1-e)(1-3e)K(e)}{E(e)-(1-e)K(e)},\\
  & f_2(e)=-\frac{(1-2e)E(e)-(1-e)K(e)}{E(e)-(1-e)K(e)}.
  \end{split}
\end{equation}
Plots of these functions are shown in Fig.~\ref{fig5} and their values at $e=0$ 
and $e=1$ are equal to
\begin{equation}\label{eq64}
\begin{split}
  &f_1(0)=f_2(0)=3,\\
  &f_1(1)=-1,\quad f_2(1)=1.
  \end{split}
\end{equation}
It is convenient to express all the variables at fixed value of time $t$ as functions of the
parameter $e$. We find  from Eq.~(\ref{eq62})
\begin{equation}\label{eq65}
  \begin{split}
   2t=2f_1\e^{2y}+f_2\e^{4y},\quad
   2x=-2f_1\e^{2y}+f_2\e^{4y},
  \end{split}
\end{equation}
so that the first equation gives
\begin{equation}\label{eq66}
  \e^{2y}\equiv\frac{1+v}{1-v}=\sqrt{\left(\frac{f_1}{f_2}\right)^2+\frac{2t}{f_2}}
  -\frac{f_1}{f_2},
\end{equation}
where we have chosen the positive root because of positivity of the function  $\e^{2y}$
for all $t$. Substitution of $\e^{2y}$ into the second formula (\ref{eq65}) gives
\begin{equation}\label{eq67}
  x=x(e)=t-2f_1\left(\sqrt{\left(\frac{f_1}{f_2}\right)^2+\frac{2t}{f_2}}
  -\frac{f_1}{f_2}\right).
\end{equation}
This formula defines the dependence of $e$ on $x$ at the fixed moment of time $t$.
The dependence of the velocity $v$ on $e$ we find from Eq.~(\ref{eq56}):
\begin{equation}\label{eq68}
  v=v(e)=\frac{\sqrt{f_1^2-2f_2t}-f_1-f_2}{\sqrt{f_1^2-2f_2t}-f_1+f_2}.
\end{equation}
This formula together with Eq.~(\ref{eq67}) define the distribution of $v$ on $x$ in a parametric form.
The obtained here formulas give a particular solution of the Whitham equations about the
evolution of a nonlinear wave pattern in the sine-Gordon equation theory. To understand its
physical meaning, let us discuss the motion of its edges.

\begin{figure}
\includegraphics[width=8cm]{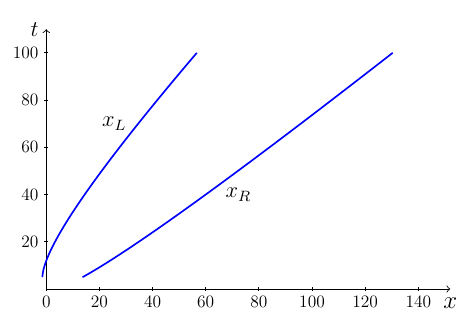}
\caption{Paths of the left $x_L$ and right $x_R$ edges of the nonlinear wave packet
in the $(x,t)$-plane calculated according to Eqs.~(\ref{eq69}) and (\ref{eq71}), respectively.}
\label{fig6}
\end{figure}

If we substitute $e=0$ in Eq.~(\ref{eq67}), then we find the law of motion of the 
small-amplitude edge of the pattern:
\begin{equation}\label{eq69}
  x_L(t)=t-6\left(\sqrt{1+2t/3}-1\right).
\end{equation}
Its velocity equals to
\begin{equation}\label{eq70}
  \frac{dx_L}{dt}=1-\frac{2}{\sqrt{1+2t/3}},
\end{equation}
that is it coincides with the flow velocity (\ref{eq68}) at $e=0$: $dx_L/dt=v(0)$.

\begin{figure}
\includegraphics[width=8cm]{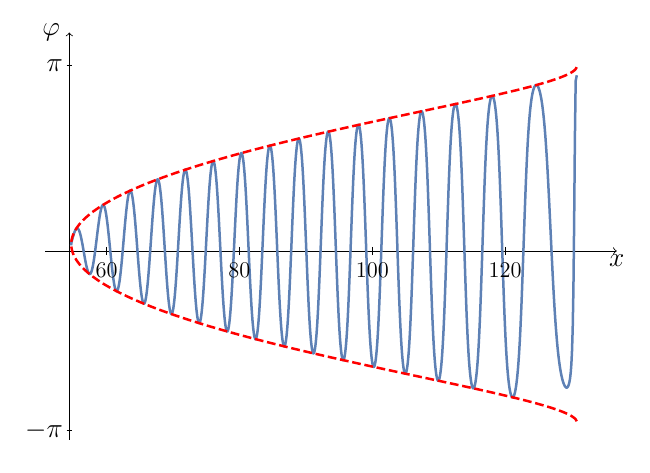}
\caption{The profile of the modulated nonlinear packet $\vphi(x,t)$ at $t=100$.
}
\label{fig7}
\end{figure}

The opposite left edge of the packet with $e=1$ moves according to the law
\begin{equation}\label{eq71}
  x_R(t)=t+2(\sqrt{1-2t}+1),
\end{equation}
and its velocity equals to
\begin{equation}\label{eq72}
  \frac{dx_R}{dt}=1+\frac{2}{\sqrt{1+2t}}.
\end{equation}
The group velocity (\ref{eq68}) at this edge has the value
\begin{equation}\label{eq73}
  v(1)=\frac{\sqrt{1+2t}}{\sqrt{1+2t}+2},
\end{equation}
that is it is equal to the phase velocity
of the wave at this point: $dx_R/dt=V=1/v(1)$. This means that at the edge with $e\to1$ 
we get a train of kinks (\ref{eq26b}) with alternating polarities. The paths of the edges 
of the wave packet are shown in Fig.~\ref{fig6}. It is clear that the pattern expands with 
the growth of time, and its soliton front propagates into the unstable state of the system. 
A typical profile is shown in Fig.~\ref{fig7}. It shows the wave propagating to the right 
into the unstable state of the system. On the left of this wave we have the stable state 
with $\vphi=0$. This instability front wave is qualitatively similar to the right 
instability front described by the self-similar solution (\ref{eq36}). However, it is 
clear that it corresponds to some initial condition with a sharp transition from the 
state with $\vphi=0$ at the left edge to the unstable state with $\vphi=\pi$ on the right 
edge rather than a localized disturbance of the unstable state.

\section{Conclusion}

In case of the focusing NLS equation, the velocity of the instability front is equal to the 
maximal group velocity of linear waves described by the stable mode outside the instability 
range of wave numbers (see \cite{kamch-92,egkk-93,bk-94}). In case of the sine-Gordon 
equation theory, the situation is very different. Now there is no such stable mode of 
linear waves, and the maximal group velocity does not exist. At the same time, the 
modulations of periodic solutions can propagate with velocities in the range $0<v<1$. 
As a result, the instability front propagates at asymptotically large times with velocity 
tending to its maximal value $v=1$. Correspondingly, the instability front consists of a 
modulated train of kinks with alternating polarities, whereas far behind it we get a wave 
of small amplitude oscillations around the stable state. It seems quite plausible that 
such a scenario is typical for conservative modulationally unstable systems that do not 
have stable modes of linear waves with maximal group velocity.

\section*{Data Availability Statement}

Data available on request from the author.

\section*{Author Declaration Section}

The author has no conflicts to disclose.


\begin{thebibliography}{99}

\bibitem{scott-07} A. Scott, {\it Nonlinear Science. Emergence and Dynamics of Coherent Structures,}
Oxford University Press (2003)

\bibitem{kpp-37} A. Kolmogoroff, I. Petrovsky, N. Piscounoff, Study of the diffusion equation with growth of the quantity of matter
and its application to a biology problem, Bulletin de l'universit\'{e} d'\'{e}tat \`{a} Moscou, Ser. int., Section A, Vol. 1
(1937)

\bibitem{dl-83} G. Dee and J. S. Langer,  Propagating Pattern Selection,
Phys. Rev. Lett., {\bf 50,} 383 (1983).

\bibitem{kn-87} V. G. Kamenskii and S. V. Manakov,  Formation of stability regions from unstable states in dissipative nonlinear systems,
JETP Letters, \textbf{45}, 638-642 (1987)

\bibitem{cgk-94} A. I. Chernykh, I. R. Gabitov, and E. A. Kuznetsov,  {\it Defects of
one-dimensional vortex lattices}, in: ``Singular limits
of dispersive waves (Lyon, 1991)'', NATO Adv. Sci. Inst. Ser. B Phys.,
315-328, Plenum, New York, 1994.

\bibitem{vanSaar-03} W. van Saarloos,  Front propagation into unstable states,
Phys. Rep., {\bf 386,} 29 (2003).

\bibitem{karpman-67} V. I. Karpman, Self-modulation of nonlinear plane waves in dispersive media,
JETP Letters, \textbf{6}, 277-279 (1967).

\bibitem{kk-68} V. I. Karpman and E. M. Krushkal', Modulated waves in nonlinear dispersive media,
Sov. Phys. JETP \textbf{28}, 277-281 (1968).

\bibitem{karpman}  V. I. Karpman, {\it Nonlinear Waves in Dispersive Media,} Pergamon Press, 1974.

\bibitem{whitham-65}  G. B. Whitham, { Non-linear dispersive waves,}
Proc. Roy. Soc. Lond. A, {\bf 283,} 238 (1965).

\bibitem{whitham} G. B. Whitham, {\it Linear and Nonlinear Waves,} (New York, Wiley, 1974).

\bibitem{kamch-92} A. M. Kamchatnov, Periodic solutions and Whitham equations
for the Heisenberg continuous classical spin model,
Phys. Lett. A, {\bf 162,} 389 (1992).

\bibitem{egkk-93} G. A. El, A. V. Gurevich, V. V. Khodorovskii, and L. A. Krylov,
Modulational instability and formation of a nonlinear oscillatory structure in a ``focusing'' medium,
Phys. Lett. A, {\bf 177,} 357 (1993).

\bibitem{bk-94} R. F. Bikbaev and V. R. Kudashev,  Example of shock waves in unstable media: the
focusing nonlinear Schr\"{o}dinger equation,
Phys. Lett. A, {\bf 190,} 255 (1994).

\bibitem{kt-88} E. A. Kuznetsov and S. K. Turitsyn, Instability and collapse of solitons in media with a defocusing nonlinearity,
Zh. Eksp. Teor. Fiz. {\bf 94,} 119 (1988) [Sov. Phys. JETP, {\bf 67,} 1583 (1988)].

\bibitem{egk-06} G. A. El, A. Gammal, and A. M. Kamchatnov, Oblique Dark Solitons in Supersonic Flow of a Bose-Einstein Condensate,
Phys. Rev. Lett., {\bf 97,} 180405 (2006).

\bibitem{kp-08} A. M. Kamchatnov and L. P. Pitaevskii,  Stabilization of Solitons Generated by a Supersonic
Flow of Bose-Einstein Condensate Past an Obstacle,
Phys. Rev. Lett., {\bf 100,} 160402 (2008).

\bibitem{kk-11} A. M. Kamchatnov and S. V. Korneev,  Condition for convective instability of dark solitons,
Phys. Lett. A, {\bf 375,} 2577 (2011).

\bibitem{amo-11} A. Amo, S. Pigeon, D. Sanvitto, V. G. Sala, R. Hivet, I. Carusotto, F. Pisanello,
G. Lem\'{e}nager, R. Houdr\'{e}, E. Giacobino, C. Ciuti, A. Bramati,
 Polariton superfluids reveal quantum hydrodynamic solitons,
Science, {\bf 332,} 1167 (2011).

\bibitem{grosso-11} G. Grosso, G. Nardin, F. Morier-Genoud, Y. Léger, and B. Deveaud-Pl\'{e}dran,
Soliton Instabilities and Vortex Street Formation in a Polariton Quantum Fluid,
Phys. Rev. Lett. {\bf 107,} 245301 (2011).

\bibitem{akns-73} M.~J.~Ablowitz,  D.~J.~Kaup, A.~S.~Newell, H.~Segur,
Method for solving the sine-Gordon equation,
Phys. Rev. Lett. {\bf 30,}  1262 (1973).

\bibitem{takhtajan-74} L. A. Takhtadjan, Exact theory of propagation of ultrashort optical pulses in two-level systems,
Zh. Eksp. Teor. Fiz., {\bf 66,} 476 (1974)
[Sov. Phys. JETP, {\bf 39,} 228 (1874)].

\bibitem{kk-76} V. A. Kozel, V. P. Kotlyarov, Almost periodic solutions of the equation $u_{tt}-u_{xx} + \sin u = 0$.
Dokl. Akad. Nauk Ukr. SSR Ser. A, {\bf 10,} 878 (1976). %Dokl. Akad. Nauk Ukrain. SSR Ser. A 1976, no. 10, 878–881

\bibitem{ks-91} V. R. Kudashev and S. E. Sharapov, Hydrodynamic symmetries for the Whitham equations
for the sine-Gordon equation, Phys. Lett. A, {bf 160,} 559-563 (1991).

\bibitem{gn-03} P. G. Grinevich, S. P. Novikov, Topological charge of the real periodic finite-gap Sine-Gordon solutions,
Comm. Pure Appl. Math., {\bf 56,} 956 (2003).

\bibitem{cmkw} J.~Cuevas-Maraver, P.~G.~Kevrekidis, F.~Williams, (Eds.) {\it The sine-Gordon Model
and its Applications,} (Cham, Springer, 2014).

\bibitem{maslov-69} V. P. Maslov, Transition of the Heisenberg equation in the limit $\hbar\to0$ to equations of one-atomic
ideal gas and quantization of relativistic hydrodynamics, %Переход при $\hbar\to0$ уравнения Гайзенберга в уравнение динамики
%одноатомного идеального газа и квантование релятивистской гидродинамики,
Teor. Mat. Fiz., {\bf 1,} 378 (1969) [Theor. Math. Phys. {\bf 1,} 289 (1969)].

\bibitem{LL6}  L. D. Landau and E. M. Lifshitz, Course of Theoretical
Physics, Vol. 6: Fluid Mechanics (Pergamon, New York, 1987).

\bibitem{LL5}  L. D. Landau and E. M. Lifshitz, Course of Theoretical
Physics, Vol. 5: Statistical Physics, Part~1 (Pergamon, New York, 1969).

\bibitem{khal-54} I. M. Khalatnikov, Some questions of relativistic hydrodynamics, Zh. Eksp. Teor. Fiz. {\bf 27,} 529 (1954).

\bibitem{as-79}  M. Abramowitz, I. A. Stegun, {\it Handbook of Mathematical Functions,}
(Dover, New York, 1965).

\bibitem{lighthill-1965} M. J. Lighthill, Contribution to the theory of waves in non-linear dispersive systems,
J. Inst. Math. Appl. {\bf 1,}  269 (1965).

\bibitem{hayes-73} W. D. Hayes,  Group velocity and nonlinear dispersive wave propagation,
Proc. Roy. Soc. Lond. A, {\bf 332,} 199 (1973).

\bibitem{gs-70} A. V. Gurevich, A. B. Shvartsburg, Exact solutions of nonlinear geometric optics equations,
 %Точные решения уравнений нелинейной геометрической %оптики,
Zh. Eksp. Teor. Fiz., {\bf 58,} 2012 (1970)
[Sov. Phys. JETP, {\bf 31,} 1084 (1970)].

\bibitem{ww-27}  E. T. Whittaker and D. N. Watson, {\it A Course of Modern
Analysis,} (Cambridge Univ., Cambridge, 1927).










\end{thebibliography}
\end{document}